\documentclass[12pt]{article}
\usepackage{amsmath}
\def\be{\begin{equation}}
\def\ee{\end{equation}}
\def\arr{\begin{array}{rll}}
\def\ea{\end{array}}
\def\bea{\begin{eqnarray}}
\def\eea{\end{eqnarray}}

\def\N2{$N{=}2$}

\def\>{\rangle}
\def\<{\langle}
\def\+{\dagger}
\def\={\ =\ }

\begin{document}
\renewcommand{\thefootnote}{\fnsymbol{footnote}}
\begin{titlepage}
\setcounter{page}{0}
\vskip 1cm
\begin{center}
{\LARGE\bf  Ricci--flat spacetimes  admitting}\\
\vskip 0.5cm
{\LARGE\bf   higher rank Killing tensors }\\
\vskip 2cm
$
\textrm{\Large Marco Cariglia\ }^{a} ~ \textrm{\Large and \ } \textrm{\Large Anton Galajinsky\ }^{b}
$
\vskip 0.7cm
${}^{a}$
{\it DEFIS, Universidade Federal de Ouro Preto, Campus Morro do Cruzeiro, \\ 35400-000 Uoro Preto, MG, Brasil}
\vskip 0.5cm
${}^{b}$
{\it
Laboratory of Mathematical Physics, Tomsk Polytechnic University, \\
634050 Tomsk, Lenin Ave. 30, Russian Federation} \\
\vskip 0.4cm
{E-mails: marco@iceb.ufop.br,~ galajin@tpu.ru}

\vskip 0.5cm

\end{center}
\vskip 1cm
\begin{abstract} \noindent
Ricci--flat spacetimes of signature $(2,q)$ with $q=2,3,4$ are constructed which admit irreducible Killing tensors of rank--3 or rank--4.
The construction relies upon the Eisenhart lift applied to Drach's two--dimensional integrable systems which is followed by the oxidation with respect to free parameters.
In four dimensions, some of our solutions are anti--self--dual.
\end{abstract}

\vspace{0.5cm}

\end{titlepage}

\renewcommand{\thefootnote}{\arabic{footnote}}
\setcounter{footnote}0

\noindent
{\bf 1. Introduction}\\

\noindent
In Riemannian geometry, a Killing tensor of rank $n$ is defined to be a totally symmetric tensor field $K_{i_1\dots i_n}$  which obeys the differential equations $\nabla_{(i_1} K_{i_2 \dots i_{n+1} )}=0$ where $\nabla_i$ is the covariant derivative and the braces indicate symmetrization of indices. A specific feature of a Killing tensor which distinguishes it from a Killing vector is that there is no coordinate transformation associated with it which would leave metric invariant. These somewhat enigmatic objects are thus attributed to hidden symmetries of a spacetime.

It was not until the pioneering work by Carter \cite{C}, in which it was shown that the Hamilton--Jacobi equation for a massive particle moving in the Kerr--Newman spacetime was separable due to the presence of an extra quadratic integral of motion, that the importance of Killing tensors in the general relativistic context was recognized \cite{WP}. Apart from providing a clue to establishing the complete integrability of the geodesic equations \cite{C}~\footnote{For higher dimensional generalizations see, e.g., Ref. \cite{KKPF} and references therein.} and the complete separation of
variables for the corresponding Sch\"odinger equation \cite{C1}, they help to identify a spacetime in accord with Petrov's classification \cite{WP}.

Killing tensors which might be revealed by analyzing the Hamilton--Jacobi equation are always of the second rank. A natural question arises whether there exist solutions of the vacuum Einstein equations which admit
higher rank Killing tensors. In a recent work \cite{GHKW}, Lorentzian spacetimes which admit irreducible rank--3 and rank--4 Killing tensors were constructed by applying the Eisenhat lift \cite{Eis} to Goryachev--Chaplygin and Kovalevskaya's tops. This method was further applied to a variety of integrable systems with the aim to produce other examples of irreducible higher rank Killing tensors \cite{GR}--\cite{MC}. It is important to stress, however, that none of the Lorentzian spacetimes studied in \cite{GHKW},\cite{GR}--\cite{MC} solves the vacuum Einstein equations.

The conventional Eisenhart lift \cite{Eis} is an embedding of a dynamical system with $n$ degrees of freedom $x_1,\dots,x_n$ governed by a potential $U(x)$ into an $(n+2)$--dimensional Lorentzian spacetime of signature $(1,n+1)$ parameterized by the coordinates $z^A=(t,s,x_1,\dots,x_n)$ (for a recent review with numerous examples see \cite{MC}). The equations of motion of the original system are contained within the null geodesics associated with the metric
\be
d \tau^2=g_{AB}(z) d z^A d z^B=-2 U(x) d t^2+2 dt ds+d x_i d x_i.
\nonumber
\ee
Each integral of motion which is a polynomial in momenta gives rise to a Killing tensor whose rank is equal to the degree of the polynomial.
The condition that the spacetime is Ricci--flat constraints the potential to be a harmonic function
\be
\partial_i \partial_i U(x)=0.
\nonumber
\ee

Close inspection of two--dimensional integrable models possessing a cubic (or higher) integral of motion \cite{H} shows that none of them is described by a harmonic function. Thus the construction of four--dimensional spacetimes of signature $(1,3)$ which admit higher rank Killing tensors seems to be problematic within the Eisenhart approach.

In higher dimensions, one can choose the potential to be a
homogeneous harmonic polynomial of a given degree and try to construct a cubic (or higher) integral of motion by direct method \cite{H}. Unfortunately, even for the simplest case of a three--dimensional system governed by a potential which is a
homogeneous harmonic polynomial of the third degree, the construction of a cubic integral appears to be intractable.

As was mentioned above, within the conventional Eisenhart approach the equations of motion of a dynamical system in an $n$--dimensional Euclidean space are embedded into the null geodesics in an $(n+2)$--dimensional Lorenzian spacetime of signature $(1,n+1)$. Had one started with a model in a pseudo--Euclidean spacetime of signature $(p,q)$, the uplifted system would have been formulated in a Lorenzian spacetime of signature $(p+1,q+1)$. What is more important, the restriction on the potential indicated above would have been altered as well.

The goal of this work is to construct Ricci--flat spacetimes of signature $(2,q)$ with $q=2,3,4$ which admit irreducible rank--3 or rank--4 Killing tensors. This is achieved by applying the Eisenhart lift to integrable models introduced by Drach \cite{Drach}~\footnote{For a more accessible description of Drach's systems see, e.g., Ref. \cite{T}.}.

In the next section, the Eisenhart lift is applied to a two--dimensional mechanics in pseudo--Euclidean spacetime of signature $(1,1)$. Restrictions on the potential which lead to a Ricci--flat or (anti) self--dual spacetime are obtained. In Sect. 3, out of ten integrable systems introduced by Drach \cite{Drach} we pick up two which obey the constraints formulated in Sect. 2. With their aid we then build two four--dimensional spacetimes of signature $(2,2)$ each of which admitting an irreducible Killing tensor of rank three. The metrics involve three and two arbitrary parameters, respectively. The first of them turns out to be anti--self--dual.
In Sect. 4, we employ the oxidation procedure which converts free parameters of the original integrable system into momenta canonically conjugate to extra cyclic variables so as to generate higher dimensional spacetimes of signature $(2,q)$ with $q=3,4$ which admit irreducible rank--3 or rank--4 Killing tensors. In the concluding Sect. 5 we summarize our results and discuss possible further developments.

\vspace{0.5cm}

\noindent
{\bf 2. Ricci--flat spacetimes of signature $(2,2)$ via the Eisenhart lift}\\

\noindent

Consider a four--dimensional spacetime of signature $(2,2)$ which is parametrized by the coordinates $z^A=(t,s,x,y)$ and is endowed with the metric\footnote{Metrics similar to (\ref{metric}) have been considered by Plebanski in \cite{Pl} in the context of the second heavenly equation. They differ from (\ref{metric}) by the dependence of the potential on its arguments which, in our notation, would be $U(t,y)$.  }
\be\label{metric}
d \tau^2=g_{AB}(z) d z^A d z^B=-2 U(x,y) d t^2+2 dt ds+2 dx dy,
\ee
where $U(x,y)$ is an arbitrary function. Taking into account the non--vanishing components of the Christoffel symbol
\be
\Gamma^{s}_{t x}=-U_{x}, \qquad \Gamma^{s}_{t y}=-U_y, \qquad \Gamma^{x}_{t t}=U_y, \qquad \Gamma^{y}_{t t}=U_x,
\ee
where $U_x=\frac{\partial U}{\partial x}$, $U_y=\frac{\partial U}{\partial y}$,
one can readily compute the Riemann tensor
\begin{align}
&
{R^s}_{xtx}=U_{xx}, && {R^s}_{xty}=U_{xy}, && {R^s}_{ytx}=U_{xy}, && {R^s}_{yty}=U_{yy},
\nonumber
\\[2pt]
&
{R^x}_{ttx}=-U_{xy}, && {R^x}_{tty}=-U_{yy}, && {R^y}_{ttx}=-U_{xx}, && {R^y}_{tty}=-U_{xy},
\end{align}
where $U_{xx}=\frac{\partial^2 U}{\partial x^2}$, $U_{xy}=\frac{\partial^2 U}{\partial x \partial y}$, $U_{yy}=\frac{\partial^2 U}{\partial y^2}$,
and verify that the only non--zero component of the Ricci tensor reads
\be
R_{tt}=2 U_{xy}.
\ee
Thus (\ref{metric}) belongs to the class of Ricci--flat spacetimes provided $U(x,y)$ is an additive function
\be\label{cond}
U(x,y)=u(x)+v(y)
\ee
with arbitrary entries $u(x)$ and $v(y)$.

In a spacetime of signature $(2,2)$ the equation\footnote{Here $g^{ml}$ is the inverse metric, $g=\det g_{nm}$, and $\epsilon_{lrqs}$ is the totally antisymmetric symbol with $\epsilon_{tsxy}=1$.}
\be
\frac 12 \sqrt{g} g^{ml} g^{nr} \epsilon_{lrqs} {R^p}_{kmn}=\pm {R^p}_{kqs},
\ee
defines the self--dual (upper sign) and anti--self--dual (lower sign) metrics, respectively. As is well known, the (anti) self--dual spacetimes are Ricci--flat. For the metric (\ref{metric}) the self--duality condition reads
\be\label{SDC}
U_{xx}=0, \qquad U_{xy}=0
\ee
which implies that any function of the form
\be
U(x,y)=a+b x+v(y),
\ee
where $a$, $b$ are constants and $v(y)$ is an arbitrary function of its argument, generates a self--dual solution of the Einstein equations.

In a similar fashion the anti--self--duality condition gives the restrictions on the potential
\be\label{ASDC}
U_{yy}=0, \qquad U_{xy}=0,
\ee
and, hence, the combination
\be
U(x,y)=a+by+u(x),
\ee
where $a$, $b$ are constants and $u(x)$ is an arbitrary function, yields a metric which describes an anti--self--dual spacetime.

Let us consider a two--dimensional dynamical system governed by the Hamiltonian
\be\label{ds}
H=p_x p_y+U(x,y),
\ee
where $(p_x,p_y)$ are momenta canonically conjugate to $(x,y)$ which obey the conventional Poisson brackets $\{x,p_x\}=1$, $\{y,p_y\}=1$. In contrast to the conventional Newtonian mechanics, the nonstandard kinetic term in the Hamiltonian results in the interchange of arguments of the partial derivatives which enter the equations of motion
\be\label{em}
\frac{d^2 x}{d t^2}=-U_y, \qquad \frac{d^2 y}{dt^2}=-U_x.
\ee
Similarly to the Newtonian mechanics which can be incorporated within the Eisenhart lift \cite{Eis}, the system (\ref{em}) can be included into the null geodesic equations associated with the metric (\ref{metric})
\be\label{geod}
\frac{d^2 z^A}{d \tau^2}+\Gamma^A_{BC} (z) \frac{d z^B}{d \tau} \frac{d z^C}{d \tau}=0, \qquad g_{AB}(z) \frac{d z^A}{d \tau} \frac{d z^B}{d \tau}=0.
\ee
Being rewritten in components, the leftmost equation in (\ref{geod}) reproduces (\ref{em}) along with
\be\label{first}
\frac{d t}{d \tau}=c_1, \qquad \frac{d s}{d t}-2 U(x,y)=c_2,
\ee
where $c_1$ and $c_2$ are constants of integration. The condition that the geodesic is null yields
\be
\frac{dx}{dt} \frac{dy}{dt}+\frac{ds}{dt}-U(x,y)=0,
\ee
which implies that $-c_2$ can be interpreted as the value of the conserved energy $\frac{dx}{dt} \frac{dy}{dt}+U(x,y)$.

A conserved quantity of the dynamical system (\ref{em}) which is a polynomial in the momenta $p_x=\frac{d y}{d t}$, $p_y=\frac{dx}{dt}$~\footnote{Note that, as compared to the Newtonian mechanics, the momenta $p_x$ and $p_y$ are interchanged. This is to be remembered in the next section when obtaining a Killing tensor from a conserved charge of an integrable system like (\ref{ds}).}  can be uplifted to a Killing tensor of the metric (\ref{metric}).
In view of the leftmost expression in (\ref{first}), a multiplication of a conserved charge of degree $l$ in momenta by ${\left(\frac{d t}{d \tau}\right)}^l$ yields an expression of the form
$K_{A_1 \dots A_l} (z) \frac{d z^{A_1}}{d \tau}\dots \frac{d z^{A_l}}{d \tau}$ from which the Killing tensor $K_{A_1 \dots A_l} (z)$ is obtained.

\vspace{0.5cm}

\noindent
{\bf 3. Drach systems and higher rank Killing tensors}\\

\noindent

Integrable systems of the type (\ref{ds}) which possess integrals of motion cubic in momenta have been studied by Drach \cite{Drach}. The inspection of the models in \cite{Drach} shows that only two of them respect the additivity condition (\ref{cond}). The first example is given by the potential and the cubic integral of motion\footnote{This model is superintegrable. In addition to the Hamiltonian and the cubic integral exposed in (\ref{Drach1}), there is an extra quadratic integral of motion which reads
$p_y^2+2\alpha x$. It is straightforward to check that the Hamiltonian along with the quadratic and cubic integrals of motion form a functionally independent set. Since in this work we are primarily concerned with higher rank Killing tensors, in what follows we disregard the quadratic integral of motion and a second rank Killing tensor associated with it. For the discussion of other Drach's superintegrable systems see \cite{T}.}
\bea\label{Drach1}
&&
U(x, y) = \alpha ( y -\beta x) + \frac{\gamma}{\sqrt{x}}, \quad
I=p_x p_y^2+\beta p_y^3+2 \alpha x p_x+2 x p_y \left(\alpha \beta+\frac{\gamma}{2 x^{\frac 32}} \right),
\eea
where $\alpha$, $\beta$ and $\gamma$ are real constants, which give rise to the third rank Killing tensor
\be
K_{ttx}=2x\left(\alpha \beta+\frac{\gamma}{2 x^{\frac 32}} \right), \qquad K_{tty}=2 \alpha x, \qquad K_{xxy}=1, \qquad K_{xxx}=3 \beta.
\ee
In accord with the criterion (\ref{ASDC}), the metric (\ref{metric}) with $U(x,y)$ in (\ref{Drach1}) is anti--self--dual for arbitrary values of the parameters $\alpha$, $\beta$ and $\gamma$. Its isometry group is generated by four Killing vectors
\bea\label{sr}
K_1=\partial_t, \qquad K_2=\partial_s, \qquad K_3=\partial_y+\alpha t \partial_s, \qquad K_4=t\partial_y+\left(\frac{\alpha t^2}{2} -x \right) \partial_s,
\eea
which obey the structure relations\footnote{$K_1$, $K_2$, $K_3$ form a representation of the Galilei algebra in one dimension while $K_4$ can be interpreted as the generator of accelerations. In general,
the Galilei algebra enlarged by acceleration generators involves the commutator $[H,C^{(n)}_i]=n C^{(n-1)}_i$, where $H$ is the generator of time translations and $C^{(n)}_i$ form a set of $N$ vector generators, $n=0,\dots,N-1$, in $d$ spatial dimensions, $i=1,\dots,d$. $C^{(0)}_i$ and $C^{(1)}_i$ are associated with space translations and Galilei boosts, while higher values of the index $n$ correspond to accelerations (for more details see, e.g., \cite{GM}).}
\be
[K_1,K_3]=\alpha K_2, \qquad [K_1,K_4]=K_3.
\ee
Note that the vector fields $K_2$ and $K_3$ are covariantly constant.
With the use of the conventional means it can be verified that six integrals of motion associated with the Killing vectors, the Killing tensor, and the metric are functionally independent\footnote{As usual, it suffices to verify that the corresponding gradients yield linearly independent vectors.}.

The second example is provided by the pair\footnote{This model is superintegrable. In addition to the Hamiltonian and the cubic integral exposed in (\ref{sec}), there is an extra quadratic integral of motion which reads
$x p_x p_y-y p_y^2+\frac{\beta x}{\sqrt{y}}-\alpha \sqrt{x}$. It is straightforward to verify that the Hamiltonian along with the quadratic and cubic integrals of motion form a functionally independent set. Since in this work we are primarily concerned with higher rank Killing tensors, in what follows we disregard the quadratic integral of motion and a second rank Killing tensor associated with it.}
\be\label{sec}
U(x,y)=\frac{\alpha}{\sqrt{x}}+\frac{\beta}{\sqrt{y}}, \qquad I=x p^2_x p_y-y p_x p^2_y
+ \frac{\beta x}{\sqrt{y}} p_x-\frac{\alpha y}{\sqrt{x}} p_y,
\ee
where $\alpha$ and $\beta$ are real constants, which gives rise to the third rank Killing tensor
\be
K_{ttx}=-\frac{\alpha y}{\sqrt{x}}, \qquad K_{tty}=\frac{\beta x}{\sqrt{y}}, \qquad K_{xyy}=x, \qquad K_{xxy}=-y.
\ee
The isometry group of the metric (\ref{metric}) with $U(x,y)$ in (\ref{sec}) includes two Killing vectors which prove to coincide with $K_1$ and $K_2$ in (\ref{sr}). $K_2$ is covariantly constant.

Further specification occurs if one of the parameters $\alpha$ or $\beta$ in (\ref{sec}) is zero. In agreement with the criteria (\ref{SDC}) and (\ref{ASDC}), for $\alpha=0$ the spacetime is self--dual, while for $\beta=0$ it is
anti--self--dual. Unfortunately, in this case the Killing tensor becomes reducible because the isometry group is extended by extra Killing vectors.
Choosing for definiteness $\beta=0$, one finds that, along with $K_1$ and $K_2$ in (\ref{sr}), the vector fields
\begin{align}
&
K_3=\partial_y, && K_4=t\partial_t -4 y \partial_y+4 x \partial_x-s\partial_s,
\nonumber
\\[2pt]
&
K_5=t\partial_y-x\partial_s, && K_6=x\partial_t-s\partial_y+4\alpha \sqrt{x} \partial_s,
\end{align}
generate the isometry group of the metric. $K_2$ and $K_3$ prove to be covariantly constant.
All together the Killing vector fields form the algebra
\begin{align}
&
[K_1,K_4]=K_1, &&  [K_1,K_5]=K_3, &&  [K_2,K_4]=-K_2, && [K_2,K_6]=-K_3,
\nonumber
\\[2pt]
&
[K_3,K_4]=-4 K_3, && [K_4,K_5]=5 K_5, && [K_4,K_6]=3 K_6. &&
\end{align}
The simplest way to demonstrate that the Killing tensor is reducible is to consider seven integrals of motion for the geodesic equations which correspond to six Killing vectors exposed above and the condition that the geodesic is null and to verify that they are functionally independent. Adding one more integral of motion associated with the Killing tensor yields a functionally dependent set which correlates with the fact that one can have at most seven functionally independent integrals of motion for a system with four degrees of freedom.

Concluding this section, we note that, while the metrics above are singular on the hyperplanes $x=0$ and $x=0$, $y=0$, respectively, the corresponding Kretschmann scalar $R_{klmn} R^{klmn}$ proves to be identically zero. This is typical for plane wave solutions. Yet, the spacetimes are not geodesically complete as geodesics emanating from a given point cannot be extended to infinite values of the affine parameter in both directions.

\vspace{0.5cm}

\noindent
{\bf 4. Higher dimensional solutions via the oxidation}\\

\noindent

Whenever an integrable system involves free parameters, it can be used to construct new integrable models which contain more degrees of freedom. This is done by converting each free parameter into a momentum conjugate to an extra cyclic variable and adding the conventional kinetic term associated with the new canonical pair to the Hamiltonian of the original system. In general, the spacetime associated with such a dynamical system does not solve the vacuum Einstein equations. In this section, we build $d=5$ and $d=6$ Ricci--flat spacetimes which admit Killing tensors of rank either three or four.

Our first example is a five--dimensional spacetime of signature $(2,3)$ which admits a rank--4 Killing tensor. It is obtained from
the model (\ref{Drach1}) by implementing the oxidation with respect to the parameter $\beta$
\be\label{Drach3}
H=\frac 12 p_w^2+p_x p_y-\alpha x p_w+\alpha y+\frac{\gamma}{\sqrt{x}}, \qquad I=p_w p_y^3+p_x p_y^2+2 \alpha x p_x+2 x p_y \left(\alpha p_w+\frac{\gamma}{2 x^{\frac 32}} \right).
\ee
Here $(w,p_w)$ is a new canonical pair obeying the conventional Poisson bracket $\{w,p_w\}=1$. Because the variable $w$ is cyclic, the function $I$ Poisson commutes with the Hamiltonian and thus provides an integral of motion quartic in momenta.

Applying the Legendre transform with respect to the momenta $(p_x,p_y,p_w)$, one obtains the Lagrangian
\be
\mathcal{L}=\frac 12 {\dot w}^2+\dot x\dot y+\alpha x \dot w-\alpha y-\frac{\gamma}{\sqrt{x}}+\frac{{(\alpha x)}^2}{2},
\ee
from which the Eisenhart metric is obtained\footnote{Given the Lagrangian $L$, the Eisenhart metric reads $d \tau^2=2 L dt^2+2 dt ds$, where $s$ is a new variable \cite{Eis}.}
\be\label{gen}
d \tau^2=-2 \left(\alpha y+\frac{\gamma}{\sqrt{x}}-\frac{{(\alpha x)}^2}{2} \right) dt^2+2 dt ds+2 dx dy+2 \alpha x dt dw+dw^2.
\ee
Because the effective potential which enters the metric
\be
U_{eff}(x,y)=\alpha y+\frac{\gamma}{\sqrt{x}}-\frac{{(\alpha x)}^2}{2}
\ee
is an additive function with respect to the arguments $x$ and $y$, the metric (\ref{gen}) solves the vacuum Einstein equations.

Taking into account the relations which link the momenta and velocities
\be
p_x=\frac{dy}{dt}, \qquad p_y=\frac{dx}{dt}, \qquad p_w=\frac{dw}{dt}+\alpha x,
\ee
one can finally obtain from $I$ in Eq. (\ref{Drach3}) the fourth rank Killing tensor which is defined on the spacetime endowed with the metric (\ref{gen})
\begin{align}
&
K_{tttx}=\frac{{(\alpha x)}^2}{2}+\frac{\gamma}{4\sqrt{x}}, && K_{ttty}=\frac{\alpha x}{2}, && K_{ttxw}=\frac{\alpha x}{6},
\nonumber
\\[2pt]
&
K_{txxx}=\frac{\alpha x}{4}, && K_{txxy}=\frac{1}{12}, && K_{xxxw}=\frac 14.
\end{align}
Other examples are constructed in a similar fashion. Omitting the details, we display below metrics in $d=5,6$ and the corresponding Killing tensors.

The oxidation of the model (\ref{Drach3}) with respect to the parameter $\gamma$ yields the $d=6$ metric
\be\label{gen1}
d \tau^2=-2 \left(\alpha y-\frac{1}{2x}-\frac{{(\alpha x)}^2}{2} \right) dt^2+2 dt ds+2 dx dy+2 \alpha x dt dw-\frac{2}{\sqrt{x}} dt du+dw^2+du^2,
\ee
where $u$ is a new coordinate, which admits the rank--4 Killing tensor
\begin{align}
&
K_{tttx}=\frac{{(\alpha x)}^2}{2}-\frac{1}{4x}, && K_{ttty}=\frac{\alpha x}{2}, && K_{ttxw}=\frac{\alpha x}{6}, && K_{txxx}=\frac{\alpha x}{4},
\nonumber
\\[2pt]
&
K_{txxy}=\frac{1}{12}, && K_{xxxw}=\frac 14, && K_{ttxu}=\frac{1}{12 \sqrt{x}}. &&
\end{align}

Note that a similar oxidation of the Hamiltonian (\ref{Drach3}) with respect to the parameter $\alpha$ yields the effective potential
\be
U_{eff}(x,y)=\frac{\gamma}{\sqrt{x}}-\frac{y^2}{2(1-x^2)},
\ee
which is not an additive function in $x$ and $y$. The corresponding spacetime thus does not solve the vacuum Einstein equations.

Further five--dimensional examples are obtained from the integrable model (\ref{sec}) by implementing the oxidation with respect to either $\alpha$ or $\beta$. Because the system holds invariant under the interchange $x\leftrightarrow y$, $\alpha\leftrightarrow \beta$, it suffices to consider $\alpha$. This gives the metric
\be\label{gen2}
d \tau^2=-2 \left(\frac{\beta}{\sqrt{y}} -\frac{1}{2x}\right) dt^2+2 dt ds+2 dx dy-\frac{2}{\sqrt{x}} dt dw+dw^2,
\ee
where $w$ is a new coordinate which along with $(t,s,x,y)$ parametrizes the spacetime, and the third rank Killing tensor
\be
K_{ttx}=\frac{y}{x}, \qquad K_{tty}=\frac{\beta x}{\sqrt{y}}, \qquad K_{xyy}=x, \qquad K_{xxy}=-y, \qquad K_{txw}=-\frac{y}{2 \sqrt{x}}.
\ee
The effective potential, which enters the braces in (\ref{gen2}), is an additive function which implies that the metric solves the vacuum Einstein equations.

Our last example is a six-dimensional spacetime which admits a third rank Killing tensor. It is obtained from (\ref{sec}) by the oxidation of both $\alpha$ and $\beta$. Denoting the extra coordinates by $w$ and $u$, one gets
the metric
\be\label{gen3}
d \tau^2=-2\left(-\frac{1}{2x} -\frac{1}{2y}\right) dt^2+2 dt ds+2 dx dy-\frac{2}{\sqrt{x}} dt dw-\frac{2}{\sqrt{y}} dt du+dw^2+du^2,
\ee
while, upon the substitution $\alpha~\rightarrow ~p_w$, $\beta~\rightarrow~p_u$, the cubic integral of motion $I$ in (\ref{sec}) yields
\begin{align}
&
K_{ttx}=\frac{y}{x}, && K_{tty}=-\frac{x}{y}, && K_{txw}=-\frac{y}{2 \sqrt{x}},
\nonumber
\\[2pt]
&
K_{tyu}=\frac{x}{2\sqrt{y}}, && K_{xxy}=-y, && K_{xyy}=x.
\end{align}
Similar to the examples above, the effective potential $U_{eff}(x,y)=-\frac{1}{2x} -\frac{1}{2y}$ is an additive function which guarantees that the metric (\ref{gen3}) is a solution of the $d=6$ vacuum Einstein equations.

Finally, one can verify that the Kretschmann scalar vanishes everywhere for the solutions above. Similar to the four--dimensional examples, the higher--dimensional generalizations are not geodesically complete
because geodesics emanating from a given point cannot be extended to infinite values of the affine parameter in both directions.

\vspace{0.5cm}

\noindent
{\bf 5. Conclusion}\\

\noindent
To summarize, in this work we applied the Eisenhart lift to Drach's integrable models and built Ricci--flat spacetimes of signature $(2,q)$ with $q=2,3,4$ which admit irreducible rank--3 or rank--4 Killing tensors. It was demonstrated that a rank--3 Killing tensor can be constructed in $d=4,5,6$, while in $d=5,6$ one can built a rank--4 Killing tensor. For the four--dimensional examples the full isometry group was found. One of our $d=4$ metrics was shown to be anti--self--dual.

While our solutions are apparently unphysical because of their signature, they do provide nontrivial specimens of Ricci--flat spacetimes admitting higher rank Killing tensors. To the best of our knowledge, examples of such a kind have not been reported in the literature yet.

The research presented in this work can be continued in several directions. First of all, it would be interesting to study whether some of the examples constructed in this work can be extended so as to form geodesically complete spacetimes. Secondly, one can try to systematically extend Drach's work with the aim to construct quartic or higher integrals of motion by implementing the direct approach \cite{H}. These would lead to new nontrivial examples of Ricci--flat spacetimes of signature $(2,q)$ in $d\geq 4$ with Killing tensors of the fourth rank or higher. Thirdly,
from a physical point of view an important problem to study is whether an integrable system with a cubic (or higher) integral of motion can be constructed which is governed by a harmonic potential. In this case Ricci--flat spacetimes of Lorenzian signature $(1,q)$ and $d>4$ admitting higher rank Killing tensors could be immediately built by applying the Eisenhart lift. Finally, it would be interesting to understand whether our solutions may find some applications within the context of the $N=2$ string \cite{BGPPR} and systems of ODEs which give rise to anti--self--dual structures \cite{CDT}.
\vspace{0.5cm}

\noindent{\bf Acknowledgements}\\

\noindent
We should like to thank Maciej Dunajski and Gary Gibbons for reading the manuscript, comments and suggestions.
The work of A.G. was supported by the MSE program Nauka under
the project 3.825.2014/K and the TPU grant LRU.FTI.123.2014.

\end{document}